\documentclass[twocolumn]{aastex701}

\usepackage{amsmath}
\usepackage{amssymb}
\usepackage{booktabs}
\usepackage{listings}

\newcommand{\resnettransformer}{ResNet--Transformer}
\newcommand{\resnet}{ResNet50}

\shorttitle{Hybrid CNN--Transformer Solar Flare Prediction}
\shortauthors{Cao \& Zhao}

\journalinfo{}

\graphicspath{{./}{figures/}}

\begin{document}

\title{Solar Flare Prediction Using a Hybrid Convolutional Neural Network and Transformer Model}

\author[orcid=0009-0004-0817-0271, gname=Mason, sname=Cao]{Mason Cao}
\affiliation{Independent Researcher, California, USA}
\altaffiliation{Present address: University of California, Santa Barbara, Santa Barbara, CA 93106, USA}
\email{masoncao1@gmail.com}

\author[orcid=0000-0002-6308-872X, gname=Junwei, sname=Zhao]{Junwei Zhao}
\affiliation{W. W. Hansen Experimental Physics Laboratory, Stanford University, Stanford, CA 94305-4085, USA}
\email{junwei@sun.stanford.edu}

\begin{abstract}
Solar flares are intense bursts of electromagnetic radiation that occur when
stored magnetic energy in the Sun's atmosphere is suddenly released. They are
categorized into five classes---from least to most powerful: A, B, C, M, and
X---with each successive class representing a ten-fold increase in energy
output. The electromagnetic radiation emitted by the stronger classes (C, M,
and X) is capable of causing significant disruptions to communication systems,
satellites, and power grids on Earth. Accurate prediction of solar flares is
crucial for mitigating their adverse effects and ensuring the functionality of
critical infrastructure. This research introduces a novel model named
\resnettransformer, which combines a convolutional neural network (CNN),
\resnet, with a standard Transformer architecture. The hybrid model effectively
processes both spatial and time-series data derived from solar images to
predict the occurrence, class (C, M, and X), and probability of solar flares
within 24\,hr, 36\,hr, and 72\,hr windows. This hybrid deep learning model
represents the first of its kind in the domain of image-based, multiclass solar
flare prediction. We evaluated the model's performance using a comprehensive
set of metrics, including weighted precision, recall, and $F_1$ score, together
with balanced accuracy, the Matthews correlation coefficient (MCC), Cohen's
$\kappa$, and the area under the receiver operating characteristic curve
(ROC--AUC). Results show that \resnettransformer\ surpasses
traditional machine learning methods, such as support vector machines (SVMs)
and standalone CNN models, across all evaluated metrics. This study highlights
the potential of integrating convolutional neural networks with Transformers to
enhance predictive capabilities in solar physics, paving the way for more
reliable and timely solar flare forecasting.
\end{abstract}


\section{Introduction}
\label{sec:intro}

Solar flares are intense bursts of radiation originating from the Sun's
atmosphere, primarily occurring within active regions (ARs) characterized by
strong magnetic fields and sunspot formations. These flares, classified into
five categories---A, B, C, M, and X---vary widely in intensity, with X-class
flares representing the most intense. The resulting phenomena, often
accompanied by coronal mass ejections and solar energetic particles, can have
significant impacts on Earth's technological infrastructure, such as
satellites, power grids, and communication systems
\citep{qahwaji2007,bobra2016,liu2017,abduallah2022,moreland2022}. Hence, it is
essential to establish a reliable model for solar flare prediction to minimize
the hazards caused by powerful flares
\citep{daglis2004,liu2019,zhang2022}.

Previous studies have designed statistical models to predict flares based on
the physical properties of active regions
\citep{gallagher2002,leka2007,mason2010}. However, physics-based models remain
far from satisfactory for accurately predicting flares ahead of time. The
availability of large-scale datasets has driven the adoption of more
sophisticated machine learning methods such as support vector machines (SVMs)
and convolutional neural networks (CNNs) for flare forecasting
\citep{bobra2015,liu2020,abduallah2021}. The majority of recent studies have
focused on developing binary classifiers based on magnetic parameters taken
from Space-weather HMI Active Region Patches (SHARP) and related data products
\citep{bobra2014}.

Although solar images and multiclass classifiers have rarely been studied,
owing to their complexity and the need for intensive computation, a few
researchers have addressed this topic. \citet{sun2022} employed
three-dimensional (3D) CNNs to forecast $\geq$M-class and $\geq$C-class flares
using SHARP magnetogram image data, and \citet{zheng2019} developed a modified
CNN model for multiclass flare prediction; see also \citet{li2020} and
\citet{georgoulis2021}. While these studies have shown promising results, their
reliance on static image data leaves opportunities for further refinement.

A few recent studies have begun to investigate how to process temporal data for
more accurate predictions. One notable example is SolarFlareNet
\citep{abduallah2023}, which employs a Transformer-based framework to analyze
the temporal dynamics of solar magnetic parameters. SolarFlareNet is designed
for operationally forecasting solar flares within 24\,hr, 48\,hr, and 72\,hr
windows. It builds three binary classifiers for the $\geq$M5.0 class, the
$\geq$M class, and the $\geq$C class; these classifiers are then integrated to
predict both the flare class and its associated probability. However,
SolarFlareNet relies solely on magnetic parameters as inputs and employs binary
classifiers, leaving room for further enhancement. Specifically, developing a
multiclass classifier with solar image data could further improve forecasting
accuracy and reliability.

Our research aims to implement a comprehensive solution by processing solar
image data and performing multiclass prediction, thereby achieving enhanced
accuracy and reliability. By leveraging a fine-tuned two-dimensional CNN,
specifically \resnet\ \citep{he2016}, to capture intricate spatial patterns in
solar images of active regions, and a Transformer architecture to model
time-dependent patterns, our hybrid model, \resnettransformer, offers improved
prediction accuracy through the Transformer's attention mechanisms.
Furthermore, our study uses recent solar data from 2023--2024, ensuring the
model is trained on up-to-date information reflective of current solar activity
trends.

\section{Experimental Design}
\label{sec:design}

\subsection{Rationale of Methodology}
\label{sec:rationale}

Solar images are seldom utilized for solar flare prediction, primarily because
of their inherent complexity. Full-disk solar images encompass multiple active
regions and various sources of noise, making it challenging to extract
high-quality data suitable for machine learning models. Consequently, the
research community has predominantly relied on SHARP magnetic parameters, as
evidenced by most prior studies. However, solar images contain valuable
information---both spatial and temporal---that extends beyond what is captured
by predefined magnetic parameters, thereby offering substantial potential for
improving flare prediction accuracy.

\begin{figure*}[ht!]
\centering
\includegraphics[width=\textwidth]{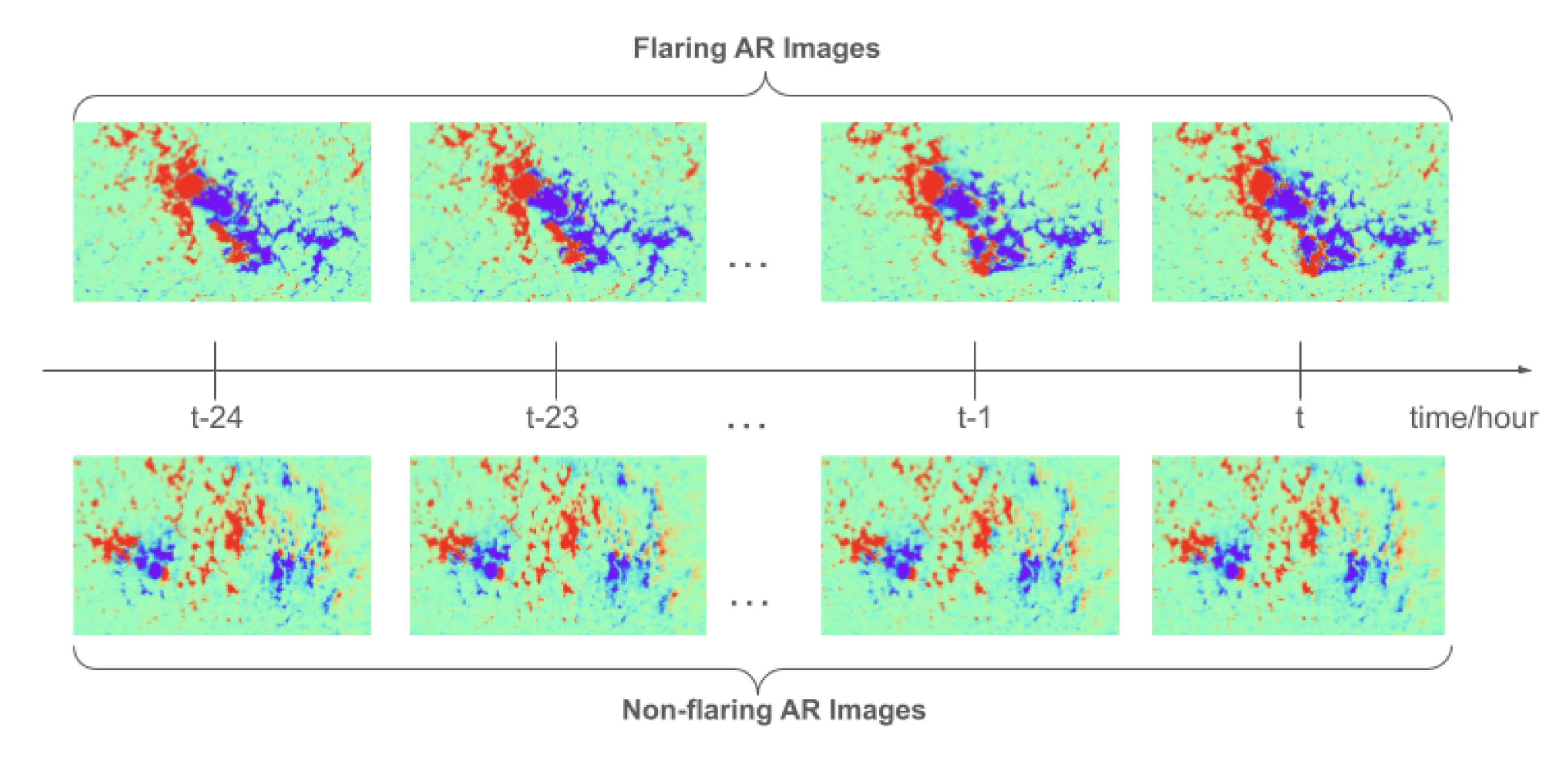}
\caption{Temporal evolution of flaring AR images (AR 13181) for an M-class
flare (top) and non-flaring AR images (AR 13177; bottom), from 24\,hr prior to
the flare's occurrence up to the moment it occurs at time $t$. Images were
generated using \textsf{SunPy} with SHARP data from 2023 January 9.}
\label{fig:ar_evolution}
\end{figure*}

Recognizing that solar flares originate from specific ARs, our study
strategically focused on AR images as the primary data source. An analysis of
the spatial and temporal properties of solar ARs, as depicted in
Figure~\ref{fig:ar_evolution}, reveals significant distinctions between active
regions that produce flares and those that do not, demonstrated by variations
in intensity, area, and location. ARs identified as flaring ARs are those where
a flare occurs at a specific time $t$, whereas non-flaring ARs remain flare-free
throughout their lifecycle. By examining the evolution of AR images within a
temporal window (e.g., 24\,hr prior to flare occurrence), our study confirms
that these observable patterns are consistently present across SHARP databases,
thereby supporting the viability of utilizing AR images for predictive
modeling.

The intricate spatial and temporal patterns inherent in AR images are well
suited to a hybrid machine learning approach that integrates CNNs with
Transformer architectures, named the \resnettransformer\ model in this study.
\resnettransformer\ combines \resnet, a 50-layer deep CNN architecture renowned
for its performance in image classification tasks, with a standard Transformer
model, known for its proficiency in sequence modeling \citep{vaswani2017}. In
our research, \resnet\ was fine-tuned to extract meaningful spatial features
from both flaring and non-flaring AR images, leveraging its deep residual
connections to capture the complex morphological characteristics essential for
identifying flare precursors. Complementing this, the Transformer architecture
was integrated to model the temporal dynamics and sequential dependencies
characteristic of solar flare activity. The Transformer's self-attention
mechanisms excel in discerning long-range temporal correlations, facilitating
the identification of temporal patterns that precede flare events. A
comprehensive data collection and labeling procedure was employed for training
the \resnettransformer\ model, which is detailed in the subsequent sections.

\subsection{Workflow}
\label{sec:workflow}

\begin{figure*}[ht!]
\centering
\includegraphics[width=0.95\textwidth]{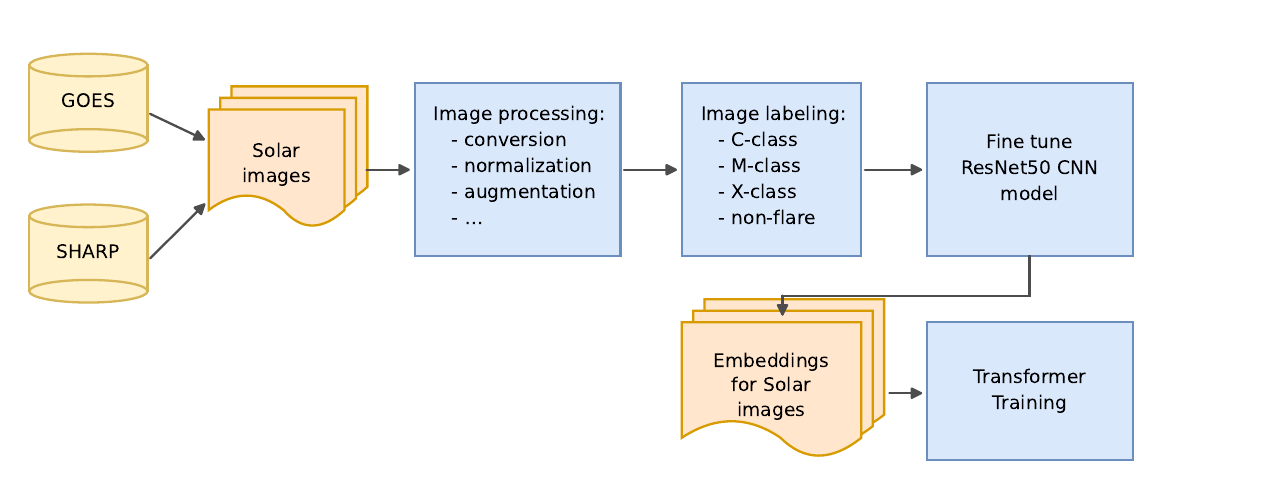}
\caption{The overall workflow for the hybrid \resnettransformer\ model training
with solar images.}
\label{fig:workflow}
\end{figure*}

Figure~\ref{fig:workflow} demonstrates the overall workflow of how the
\resnettransformer\ model was fine-tuned for use in solar flare prediction.

In this workflow, solar image data in ARs are first collected by querying the
SHARP and GOES databases; they are then processed and labeled before being fed
into a pre-trained \resnet\ CNN model. After fine-tuning, \resnet\ transforms
the image data into embeddings, with each embedding being a vector of 2048
numerical values that represents a corresponding image. Since the Transformer
is unable to process image data directly, it requires a sequence of embedding
groups as input. Each embedding group consists of six consecutive embeddings,
which preserve the temporal dynamics of their six corresponding images. All
embedding groups are then fed into the Transformer for training or testing
purposes. Each component of this process is elaborated on in detail in the
subsequent sections.

\subsection{Data}
\label{sec:data}

\subsubsection{Data Collection}
\label{sec:collection}

In this study, we used SHARP magnetograms as our primary dataset. SHARP
(Space-weather HMI Active Region Patches) is a data product released by the
Solar Dynamics Observatory (SDO) science team. SHARP includes patches of vector
magnetic field data taken by the Helioseismic and Magnetic Imager (HMI)
instrument aboard SDO; these patches encapsulate automatically detected active
regions. We accessed the SHARP data with a \textsf{SunPy}
\citep{sunpy2020}-affiliated package called \textsf{drms}
\citep{glogowski2019}. These solar images (SHARP magnetograms) were sourced
from the Joint Science Operations Center (JSOC). We systematically extracted
SHARP data at hourly intervals from the \texttt{hmi.sharp\_cea\_720s} series,
ensuring a high temporal resolution to capture the dynamic nature of active
regions.

Our dataset encompassed solar flare events from 2023 January to 2024 September
(the study period). This period was chosen to incorporate the latest
advancements in solar observation technologies and to leverage recent solar
activity patterns that may influence flare dynamics. Specifically, we cataloged
C-, M-, and X-class flares based on their occurrence in the GOES (Geostationary
Operational Environmental Satellites) X-ray flare catalogs provided by the
National Oceanic and Atmospheric Administration (NOAA).

To construct a balanced and comprehensive dataset, we included both flaring and
non-flaring ARs. Flaring AR images represented positive data, where a flare
would occur in the AR. Non-flaring AR images represented negative data, where
no flare would occur in the AR during its lifetime. Data samples were gathered
for three distinct time horizons: 24\,hr, 36\,hr, and 72\,hr windows preceding
a flare event. This stratification allowed for multiclass prediction, catering
to varying operational requirements.

Figures~\ref{fig:code_flares} and \ref{fig:code_images} show snippets of the
Python scripts used for data collection (the full script is available upon
request). We first queried the GOES database through JSOC
(Figure~\ref{fig:code_flares}) to obtain a list of flaring AR numbers during
the study period. These ARs are associated with the different classes of flares
that occurred. Other metadata related to these flares, such as date, class,
start time, and end time, were also queried for the subsequent image data query
(Figure~\ref{fig:code_images}). The final result consisted of 442 C-class
flares, 127 M-class flares, and 9 X-class flares for the period from 2023
January to 2024 September.

\begin{figure}[ht!]
\begin{lstlisting}
flare_results = Fido.search(
    a.Time(start_time, end_time),
    a.hek.EventType("FL"),
    a.hek.OBS.Observatory == "GOES"
)
\end{lstlisting}
\caption{Python code for collecting flare events.}
\label{fig:code_flares}
\end{figure}

With the metadata of the flares, we selected the start time and AR number of
each flare to query SHARP magnetograms through JSOC and obtain the AR images
(Figure~\ref{fig:code_images}). We aimed to collect AR images from 24\,hr (and
likewise 36\,hr and 72\,hr) prior to the flare time, which constitute the
positive data for training our model. Accordingly, in
Figure~\ref{fig:code_images}, \texttt{this\_start\_time} is equal to the start
time of a flare minus 24\,hr, while \texttt{this\_end\_time} is the start time
of the flare. We used this process to download 24 AR images for a specific
flare, and applied the same procedure to the 36\,hr and 72\,hr windows.

\begin{figure}[ht!]
\begin{lstlisting}
flare_ar_results = Fido.search(
    a.Time(this_start_time, this_end_time),
    a.jsoc.Series("hmi.sharp_cea_720s"),
    a.Sample(1 * u.hour),
    a.jsoc.Keyword("NOAA_AR") == ar_number,
    a.jsoc.Segment("magnetogram"),
    a.jsoc.Notify("user@example.com")
)
\end{lstlisting}
\caption{Python code for collecting flaring AR images.}
\label{fig:code_images}
\end{figure}

Collecting non-flaring AR images, which served as negative data for training
our model, was not straightforward. To ensure consistency, we defined the time
range for non-flaring AR image data as $[t-23, t]$, mirroring the range used
for flaring AR images when a flare occurs at time $t$. However, obtaining the
AR numbers associated with non-flaring images proved challenging. To address
this, we employed an indirect method. First, we downloaded all AR images within
the specified time range $[t-23, t]$, creating a superset of potential
non-flaring images. We then filtered the flaring images out of this superset.
Each AR image filename contains a HARP number, which refers to an HMI Active
Region Patch. Using the HARP number, we queried the HARP-to-AR map provided by
NOAA to determine the corresponding AR numbers for all images in the superset.
After processing all AR images in this manner, we obtained a complete list of
AR numbers. We then cross-referenced these AR numbers with the GOES database
for the study period to identify any that were associated with flares. By
excluding images linked to these flaring AR numbers, the remaining images were
considered non-flaring.

Quality assurance was critical in our data collection process. Active regions
located beyond $\pm72\arcdeg$ of the central meridian were excluded to mitigate
projection effects that could distort magnetic field measurements.
Additionally, any SHARP data showing low quality, as defined by
\citet{hoeksema2014}, were systematically removed. This rigorous filtering
ensured that only high-fidelity data were utilized, enhancing the reliability
of subsequent model training and evaluation.

\subsubsection{Data Labeling}
\label{sec:labeling}

Accurate data labeling is critical to the efficacy of machine learning models,
as it establishes the ground truth against which predictions are measured. In
our framework, labeling was performed based on the temporal proximity of flare
events to the collected SHARP data samples.

For each identified solar flare classified as C-, M-, or X-class occurring at a
specific time $t$ within an AR, we designated all SHARP data samples from $t$
back to $t-24$\,hr, $t-36$\,hr, and $t-72$\,hr as positive instances for
24\,hr-, 36\,hr-, and 72\,hr-ahead predictions, respectively. This temporal
labeling approach captured the evolving magnetic conditions leading up to flare
events, providing the model with contextual information essential for accurate
forecasting.

In scenarios where multiple flares occurred in close temporal succession within
the same AR, resulting in overlapping data samples, we prioritized the labeling
based on flare intensity. Specifically, data samples overlapping multiple flare
events were labeled according to the highest-class flare present, ensuring that
the most significant event influenced the training process. Conversely, SHARP
data samples derived from non-flaring ARs were labeled as negative data,
representing periods of no imminent flare activity. This distinction between
positive and negative samples was crucial for training the model to
differentiate between conditions conducive to flaring and those that were not.

Table~\ref{tab:samples} shows the total numbers of positive and negative image
samples in each class for 24\,hr-, 36\,hr-, and 72\,hr-ahead flare occurrence.
The numbers in the table do not include low-quality data samples that were
removed according to the process described above.

\begin{deluxetable}{llrrr}
\tablecaption{Total numbers of positive and negative data samples in each class
for 24\,hr, 36\,hr, and 72\,hr prior to flare occurrences.\label{tab:samples}}
\tablehead{
\colhead{Data Period} & \colhead{Data Type} & \colhead{C Class} &
\colhead{M Class} & \colhead{X Class}
}
\startdata
24\,hr & Positive & 10,392  & 2,966   & 202 \\
       & Negative & 124,701 & 40,389  & 2,926 \\
\hline
36\,hr & Positive & 14,836  & 4,427   & 309 \\
       & Negative & 192,760 & 61,920  & 3,699 \\
\hline
72\,hr & Positive & 30,732  & 9,088   & 621 \\
       & Negative & 399,024 & 127,002 & 7,245 \\
\enddata
\end{deluxetable}

The \resnettransformer\ model requires consecutive data samples in a time
window without missing data. To address this issue, we implemented an
average-padding strategy to synthesize data samples in cases where data
continuity was disrupted by the exclusion of low-quality data. Average padding
proceeds as follows: when the missing datum was in the middle of the dataset,
we averaged the two nearest-neighbor images to create a new padding image as
the replacement for the missing datum; if there was only one neighbor (i.e.,
the missing datum was before or after the data group), the neighboring image
was used as the replacement. This approach ensured a more uniform distribution
of samples across different classes and time horizons, mitigating potential
biases during model training.

\begin{figure*}[ht!]
\centering
\includegraphics[width=0.95\textwidth]{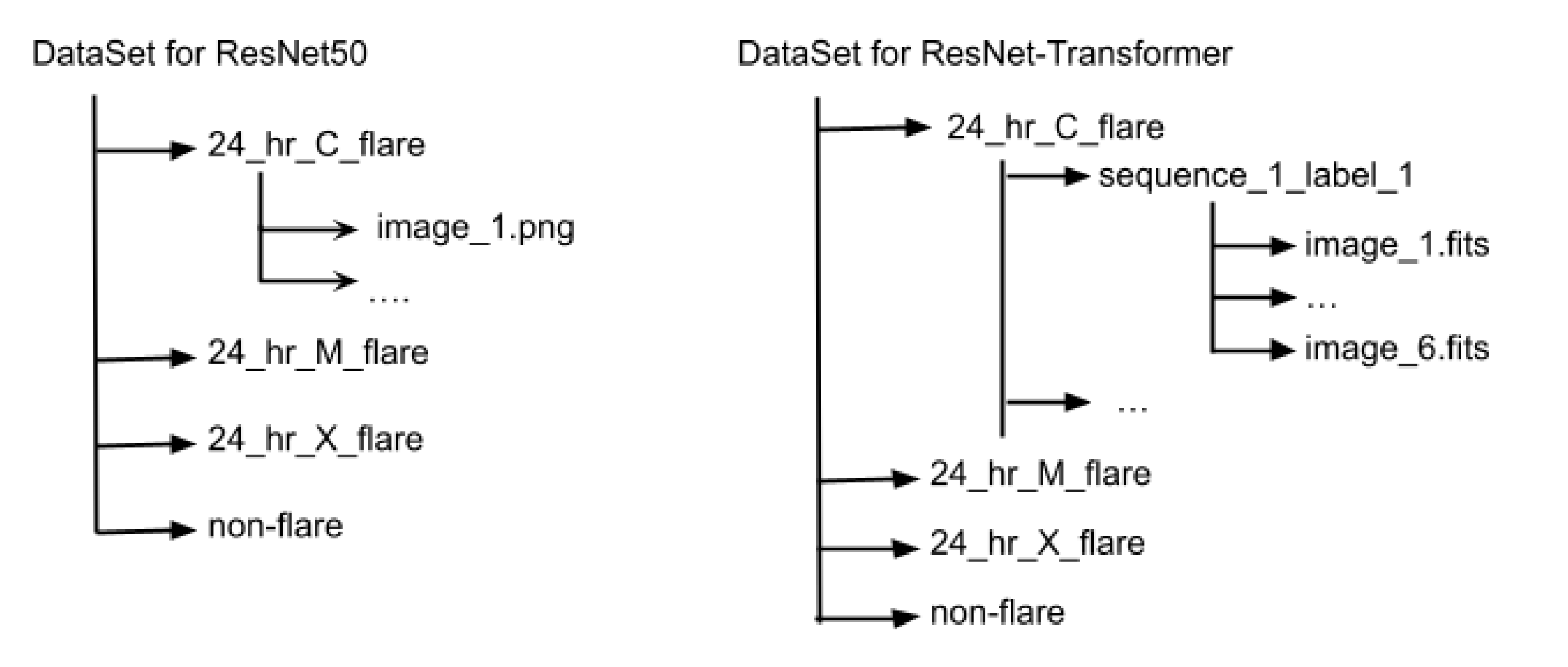}
\caption{Dataset structure for \resnet\ fine-tuning (left) and
\resnettransformer\ training (right).}
\label{fig:dataset}
\end{figure*}

We used \textsf{PyTorch}, an open-source deep learning framework, for data
loading and model training. To load data into \resnet\ for fine-tuning, we used
the \texttt{DataLoader} utility in \textsf{PyTorch} to scan image folders,
where each folder contained a list of images and the folder name was the class
name. Figure~\ref{fig:dataset} shows our image folders for the 24\,hr
experiment for both \resnet\ fine-tuning (left) and \resnettransformer\
training (right). We repeated the same process for the 36\,hr and 72\,hr
windows.

The dataset was slightly different for the hybrid \resnettransformer\ model.
Instead of using the solar AR images directly, as for \resnet, the
\resnettransformer\ model required the data to reflect temporal features, so we
created a group of images sorted by their timestamps to represent a single data
point.

We used a sliding window (with a size of six hours in this study) to create a
data group, where the label of the group (e.g., C-flare, M-flare, X-flare, or
non-flare) was determined by the class of the next image after the group. Note
that the size of the sliding window affected the performance of the hybrid
model: a larger size resulted in better performance but greatly increased the
training time. We experimented with different sizes and chose a window size of
six in our study because it provided satisfactory performance within a
reasonable training time given limited computational resources.

\begin{figure*}[ht!]
\centering
\includegraphics[width=0.9\textwidth]{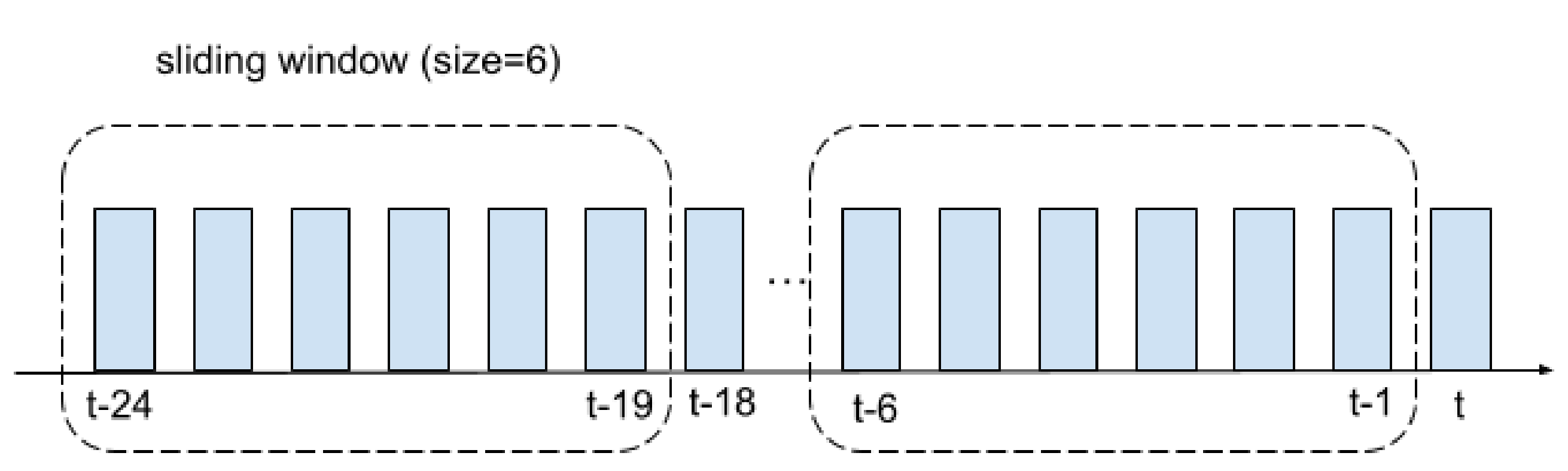}
\caption{Schema for labeling images for the \resnettransformer\ model, where a
flare occurs at time $t$.}
\label{fig:labeling}
\end{figure*}

Figure~\ref{fig:labeling} demonstrates the labeling procedure. With a flare
occurring at time $t$, we assigned the first six adjacent images from
$[t-24, t-19]$ to a group whose label was based on the AR image at $t-18$. Note
that the group $[t-6, t-1]$ was the last group, since we had no data to
indicate the group label after time $t$.

Each image group, after addressing missing data, was stored in a dedicated
folder following the naming convention
\texttt{sequence\_$\langle$index$\rangle$\_label\_$\langle$label$\rangle$},
where $\langle$index$\rangle$ was the sequence number of an image group and
$\langle$label$\rangle$ corresponded to the numerical representation of the
flare class (0 for non-flare, 1 for C-flare, 2 for M-flare, and 3 for X-flare).
The directory hierarchy is shown on the right in Figure~\ref{fig:dataset}. Each
folder contained exactly six images, consisting of real and padded images as
necessary.

To validate and ensure the robustness of our predictive models, we employed a
ten-fold cross-validation technique. The dataset was partitioned into ten
equal-sized folds; in each iteration, one fold served as the test set while the
remaining nine constituted the training set. Importantly, all data samples from
a single AR were confined to either the training or the test set within a given
fold, preventing data leakage and ensuring that the model was evaluated on
entirely unseen ARs. Performance metrics were averaged from testing results
across all ten folds, with the mean and standard deviation providing insight
into the model's consistency and reliability.

\subsubsection{Image Conversion, Normalization, Augmentation, and Data
Imbalance}
\label{sec:preprocessing}

The SHARP data obtained from JSOC were originally provided in the Flexible
Image Transport System (FITS) format, which is optimized for storing
astronomical data but is not directly compatible with \resnet. To facilitate
the use of these images in our deep learning framework, we performed a
systematic conversion process to transform the FITS images into Portable
Network Graphics (PNG) format. We then resized them to $224\times224$ pixels to
accommodate the input dimensions required by \resnet. The next crucial step
involved normalizing the pixel intensity values, which we performed by scaling
the pixel values to a standardized range, typically between 0 and 1. This
scaling was essential as it ensured uniformity across all input images,
allowing \resnet\ to learn more effectively by preventing issues related to
varying brightness and contrast levels.

To further enhance the robustness and generalization capabilities of the model,
we employed data augmentation techniques. These techniques artificially
expanded the diversity of the training dataset by applying a series of
transformations to the original images, thereby simulating various real-world
scenarios and mitigating the risk of overfitting. The augmentation strategies
implemented in this study included:
\begin{enumerate}
\item \textit{Rotation}: images were rotated by random angles within a
specified range ($15\arcdeg$ in this study) to simulate different orientations
of active regions on the solar surface.
\item \textit{Flipping}: both horizontal and vertical flips were performed to
introduce variability in the dataset.
\item \textit{Scaling}: random zooming in and out of images was applied to
capture features at multiple scales.
\item \textit{Translation}: slight shifts in the position of active regions
within the images were implemented to simulate small movements and to ensure
that the model can accurately detect flares irrespective of their exact
location within the image.
\end{enumerate}

A known issue for solar image datasets is data imbalance. From
Table~\ref{tab:samples}, the size of the negative dataset was at least 12 times
larger than that of the positive dataset for each flare class, and the data
sizes of different flare classes also varied. To mitigate this, we used two
strategies: (1) an oversampling technique to increase the size of the positive
dataset by setting a cadence of 12 minutes for flaring image sampling while
keeping a cadence of one hour for non-flaring image sampling; and (2) a
weighted-class approach to assign different weights to classes in the loss
function during model training, giving more importance to minority classes
(e.g., the X class).

These techniques enhanced the dataset's diversity and addressed data imbalance,
enabling the hybrid model to learn more generalized and robust features
essential for accurate solar flare prediction. By exposing the model to a wide
range of transformed images during training, we reduced the likelihood of
overfitting and improved its performance on unseen data.

\subsection{Methods}
\label{sec:methods}

\subsubsection{\resnet\ Fine-Tuning}
\label{sec:finetune}

We began the fine-tuning process with a pre-trained \resnet\ using weights
obtained on the general ImageNet dataset \citep{deng2009}. Recognizing the
specific requirements of solar flare prediction, we modified the network's
architecture by replacing the final fully connected (FC) layer with a new FC
layer tailored to our classification task. This new layer was designed to
output class probabilities corresponding to the presence and class of solar
flares, enabling the model to distinguish between different flare intensities.

For the training configuration, we employed the Adam optimizer
\citep{kingma2015}, an iterative optimization algorithm used to minimize the
loss function during the training of neural networks, with a learning rate of
$10^{-4}$, striking a balance between convergence speed and stability. A
weighted cross-entropy loss function was used for multiclass classification,
effectively handling the probabilistic nature of flare prediction. A batch size
of 32 was selected to ensure efficient training without exhausting
computational resources. The model was trained for 30 epochs, incorporating
early stopping to prevent overfitting by halting training when the validation
loss ceased to improve.

Our transfer learning strategy initially involved training only the modified FC
layer while keeping the convolutional layers frozen. This approach allowed the
model to adapt the high-level features specific to our task without disrupting
the pre-trained low-level feature detectors. Subsequently, the entire network
was fine-tuned with a lower learning rate, enabling the adjustment of
pre-trained weights to better suit the specific characteristics of solar
images.

During training, a validation set comprising 10\% of the training data was used
to monitor performance and apply early stopping based on improvements in the
validation loss. This strategy ensured that the model maintained its
generalization capabilities while optimizing its predictive accuracy on the
training data.

\subsubsection{Feature Extraction with the Fine-Tuned \resnet}
\label{sec:features}

After fine-tuning, the best \resnet\ model, with the highest accuracy, was
employed to extract spatial features as embeddings from solar images. These
embeddings encapsulated high-level representations of the images, facilitating
the effective temporal modeling essential for accurate solar flare prediction.

The feature extraction process began with the sequence of solar image groups. A
mask tensor was created to indicate real images (1) and padded images (0). This
mask was crucial for informing the Transformer component of the model to focus
attention on real images during training. The final classification layer of
\resnet\ was removed, retaining the convolutional layers responsible for
generating 2048-dimensional feature maps from the input images. While
maintaining the sequenced dataset structure, we also created sinusoidal
positional encodings \citep{vaswani2017} and added them to the corresponding
embeddings, because our training was executed in parallel and the positional
information of each embedding was required; otherwise, the order of the
sequence would have been lost. The ordered and positionally encoded feature
sequences were then formatted into tensors suitable for input into the
Transformer model.

\subsubsection{\resnettransformer\ Training and Testing}
\label{sec:training}

The core of the \resnettransformer\ model is a standard Transformer encoder
architecture configured with six encoder layers, each comprising eight-head
multi-head self-attention mechanisms and position-wise feedforward networks.
The embedding dimension ($d_{\mathrm{model}}$) was set to 512, and the
feedforward network dimension ($d_{\mathrm{ff}}$) was 2048, providing ample
capacity to model complex temporal relationships. A dropout rate of 0.1 was
employed to prevent overfitting, and ReLU activation functions introduced
non-linearity within the feedforward networks. The Adam optimizer was again
used, with a learning rate of $10^{-4}$ and a batch size of 32, balancing
convergence speed and computational efficiency. Training was conducted over 30
epochs, incorporating early stopping based on the validation loss to mitigate
overfitting and ensure the model's generalization capabilities.

During training, the Transformer processed chronological sequences of feature
embeddings, capturing intricate temporal patterns and dependencies indicative
of impending solar flares. The model was designed to perform multi-task
learning by simultaneously predicting the class of solar flares and estimating
the probability of flare occurrence. Specifically, the classification task
distinguished between C-, M-, and X-class flares using a weighted cross-entropy
loss function, while the probability estimation task employed a softmax
function to assess the likelihood of a flare event within a given sequence.

The held-out fold from the ten-fold cross-validation was used to test the
performance of the \resnettransformer\ model, and performance metrics were
evaluated based on the testing results.

\section{Results}
\label{sec:results}

This section presents the empirical findings of our study, evaluating the
performance of the hybrid \resnettransformer\ model and comparing it against
traditional machine learning approaches, including SVMs and a standalone CNN
model using \resnet.

\subsection{Performance of the \resnettransformer\ Model}
\label{sec:performance}

The selection of performance metrics was based on two characteristics of our
study: (1) multiclass classification, and (2) imbalanced data for the different
flare classes. Balanced accuracy and weighted versions of precision, recall,
and the $F_1$ score were chosen and adapted to accommodate multiclass
classification. Balanced accuracy accounts for class imbalance by averaging the
recall obtained on each class. Table~\ref{tab:metrics} lists the formulae for
calculating the per-class precision, recall, and $F_1$ score, along with
balanced accuracy and the weighted metrics, where TP, FP, FN, and TN stand for
true positive, false positive, false negative, and true negative, respectively.
These terms are integral components of the confusion matrix, which provides a
detailed breakdown of a model's predictions compared to the actual outcomes.
For multiclass classification, the confusion matrix is an $N\times N$ table,
where $N$ is the number of classes; the concepts of TP, FP, FN, and TN apply to
each class individually, treating it as the positive class and the rest as
negative.

\begin{deluxetable}{ll}
\tablecaption{Summary of performance metrics used in this study.
\label{tab:metrics}}
\tablehead{\colhead{Metric} & \colhead{Formula}}
\startdata
Precision (per class) &
  $\mathrm{Precision}_i = \dfrac{\mathrm{TP}_i}{\mathrm{TP}_i+\mathrm{FP}_i}$ \\[8pt]
Recall (per class) &
  $\mathrm{Recall}_i = \dfrac{\mathrm{TP}_i}{\mathrm{TP}_i+\mathrm{FN}_i}$ \\[8pt]
$F_1$ score (per class) &
  $F_{1,i} = 2\times\dfrac{\mathrm{Precision}_i\times\mathrm{Recall}_i}
  {\mathrm{Precision}_i+\mathrm{Recall}_i}$ \\[10pt]
Balanced accuracy &
  $\dfrac{1}{N}\displaystyle\sum_{i=1}^{N}\mathrm{Recall}_i$ \\[10pt]
Weighted precision &
  $\displaystyle\sum_{i=1}^{N}\left(\frac{n_i}{\sum_{j=1}^{N}n_j}
  \times\mathrm{Precision}_i\right)$ \\[10pt]
Weighted recall &
  $\displaystyle\sum_{i=1}^{N}\left(\frac{n_i}{\sum_{j=1}^{N}n_j}
  \times\mathrm{Recall}_i\right)$ \\[10pt]
Weighted $F_1$ score &
  $\displaystyle\sum_{i=1}^{N}\left(\frac{n_i}{\sum_{j=1}^{N}n_j}
  \times F_{1,i}\right)$ \\
\enddata
\tablecomments{$n_i$ is the number of true instances of class $i$ and $N$ is
the number of classes.}
\end{deluxetable}

Beyond these traditional metrics, we also analyzed the Matthews correlation
coefficient \citep[MCC;][]{gorodkin2004} and Cohen's $\kappa$
\citep{bakeman1997}, which offer more sophisticated evaluations of model
performance by accounting for the balance between true and false predictions
across all classes. The MCC is a comprehensive metric that considers true and
false positives and negatives, and is regarded as a balanced measure even when
classes are of very different sizes. Cohen's $\kappa$ measures the agreement
between two sets of predictions while accounting for agreement occurring by
chance.

The MCC for multiclass classification is calculated as
\begin{equation}
\mathrm{MCC} = \frac{c\times s-\sum_{k}^{K}p_k\times t_k}
{\sqrt{\left(s^2-\sum_{k}^{K}p_k^2\right)\times
\left(s^2-\sum_{k}^{K}t_k^2\right)}},
\label{eq:mcc}
\end{equation}
where $k$ is the class index running from 1 to $K$, $s$ is the number of
samples, $c$ is the number of samples correctly predicted, $t_k$ is the number
of times class $k$ truly occurred, and $p_k$ is the number of times class $k$
was predicted. Cohen's $\kappa$ for multiclass classification is given by
\begin{equation}
\kappa = \frac{P_0-P_e}{1-P_e},
\label{eq:kappa}
\end{equation}
where $P_0$ is the observed proportional agreement and $P_e$ is the expected
agreement by chance. In our Python code, we used the \textsf{scikit-learn}
package to calculate these metrics.

Note that we did not estimate the true skill statistic (TSS) in our study. The
TSS is defined for binary classifiers and is popular in other solar flare
studies; however, it is not directly applicable to our multiclass prediction
model.

Table~\ref{tab:performance} summarizes the mean performance metric values, with
standard deviations enclosed in parentheses, for the 24\,hr, 36\,hr, and 72\,hr
predictions made by \resnettransformer. The mean and standard deviation were
calculated from testing results across the ten folds.

Overall, \resnettransformer\ performed well across all evaluated metrics for
the 24\,hr, 36\,hr, and 72\,hr predictions, with precision above 90\% for all
three time windows. The metric values for the 36\,hr and 72\,hr windows were
lower than those for the 24\,hr window, as expected given the longer range of
those predictions.

\begin{deluxetable}{lccc}
\tablecaption{Performance metrics of the \resnettransformer\ model.
\label{tab:performance}}
\tablehead{
\colhead{Metric} & \colhead{24\,hr} & \colhead{36\,hr} & \colhead{72\,hr}
}
\startdata
Precision         & 0.9370 (0.0013) & 0.9352 (0.0082) & 0.9227 (0.0069) \\
Recall            & 0.9030 (0.0028) & 0.8972 (0.0154) & 0.8851 (0.0128) \\
$F_1$ score       & 0.9149 (0.0230) & 0.9111 (0.0116) & 0.9026 (0.0103) \\
Balanced accuracy & 0.8798 (0.0151) & 0.8702 (0.0347) & 0.8595 (0.0249) \\
MCC               & 0.7597 (0.0043) & 0.7535 (0.0280) & 0.7512 (0.0211) \\
Cohen's $\kappa$  & 0.7480 (0.0051) & 0.7403 (0.0314) & 0.7314 (0.0240) \\
\enddata
\tablecomments{Numbers represent the mean across ten cross-validation folds,
with the standard deviation in parentheses. Precision, recall, and the $F_1$
score are the weighted (multiclass) versions defined in
Table~\ref{tab:metrics}.}
\end{deluxetable}

\begin{figure}[ht!]
\centering
\includegraphics[width=\columnwidth]{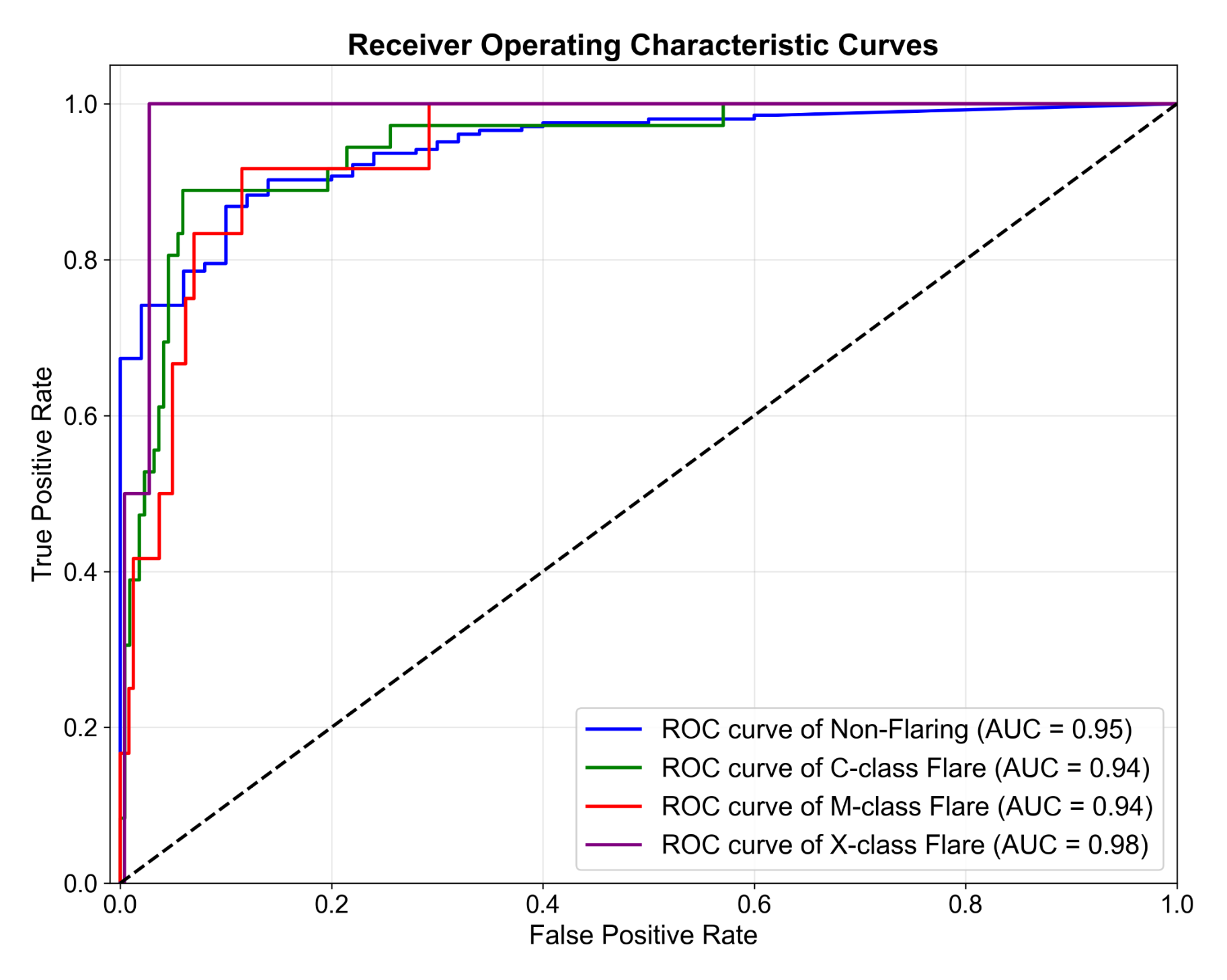}
\caption{ROC curve for each class in the 24\,hr \resnettransformer\ model. The
AUC is calculated based on one run of the testing data.}
\label{fig:roc}
\end{figure}

The area under the receiver operating characteristic curve (ROC--AUC) was
evaluated for each class through a one-vs-rest approach to assess the
discriminative power of the model. Per-class ROC curves for the 24\,hr model
are shown in Figure~\ref{fig:roc}, from which X-class flares and the
non-flaring class had the larger AUC values, followed by M-class and C-class
flares. This indicates that it is easier to distinguish X-class flares or the
non-flaring class from the remaining classes. This result is reasonable because
X-class flares have the strongest intensity while the non-flaring class has the
weakest (or no) intensity. Similar trends were seen in the 36\,hr and 72\,hr
ROC curves.

\subsection{Performance Comparison}
\label{sec:comparison}

To the best of our knowledge, this is the first time a hybrid deep learning
architecture has been used to create an image-based multiclass solar flare
prediction model, which makes it challenging to compare performance directly
with previous studies that either use magnetic parameters or build binary
classifiers. We therefore compared \resnettransformer\ with the standalone
\resnet\ CNN model. Moreover, since the SVM model was proposed by
\citet{bobra2015}, it has become a popular reference model for previous
studies; although that SVM model is trained as a binary classifier with
magnetic parameters, we still used it as a reference for comparison. The 24\,hr
model was selected for the comparison study, though similar results were
observed for the 36\,hr and 72\,hr models. Note that for the SVM model the
authors listed metrics for both positive and negative data, so we calculated
the number for each metric by averaging the positive and negative values.

The comparison results are shown in Table~\ref{tab:comparison}. The
\resnettransformer\ model outperformed both the SVM and the standalone \resnet\
CNN models across all evaluated metrics. Specifically, the \resnettransformer\
approach achieved higher precision and recall values, indicating its superior
ability to correctly identify flare events and to minimize false positives and
false negatives. The higher $F_1$ score further corroborates the model's
balanced performance, indicating that it is a more effective tool for solar
flare prediction than these traditional methods.

\begin{deluxetable}{lccc}
\tablecaption{Comparative performance of prediction models.
\label{tab:comparison}}
\tablehead{
\colhead{Model} & \colhead{Precision} & \colhead{Recall} &
\colhead{$F_1$ Score}
}
\startdata
\resnettransformer & 0.9370 & 0.9030 & 0.9149 \\
\resnet            & 0.7443 & 0.8307 & 0.7539 \\
SVM                & 0.834  & 0.805  & 0.818  \\
\enddata
\tablecomments{Values for the 24\,hr prediction window. SVM values are taken
from \citet{bobra2015} and averaged over the positive and negative classes.}
\end{deluxetable}

To determine the statistical significance of the performance difference between
\resnet\ and \resnettransformer, we conducted paired $t$-tests; the results are
summarized in Table~\ref{tab:ttest}. The results indicate that the
\resnettransformer\ model significantly outperforms the standalone \resnet\
model across all metrics considered ($p<0.05$), demonstrating its superior
capability in solar flare prediction.

\begin{deluxetable}{lll}
\tablecaption{Statistical significance of performance differences between
\resnet\ and \resnettransformer.\label{tab:ttest}}
\tablehead{
\colhead{Metric} & \colhead{$p$-value} & \colhead{Significance}
}
\startdata
Precision         & $<0.001$ & Significant \\
Recall            & $<0.001$ & Significant \\
$F_1$ score       & $<0.001$ & Significant \\
Balanced accuracy & 0.005    & Significant \\
MCC               & 0.003    & Significant \\
Cohen's $\kappa$  & 0.006    & Significant \\
\enddata
\tablecomments{A difference is deemed statistically significant if $p<0.05$
(at the 95\% confidence level).}
\end{deluxetable}

\section{Discussion}
\label{sec:discussion}

Our approach was adapted to the constraints of limited computational resources
by optimizing the dataset size and leveraging pre-trained models. Fine-tuning a
pre-trained \resnet\ model on a curated dataset maximized performance gains
while minimizing computational overhead. By utilizing multi-head self-attention,
the Transformer processed temporal sequences without incurring excessive
computational costs. Additionally, the use of gradient checkpointing, a
technique that reduces memory requirements during the training phase, further
optimized memory usage and accelerated training times, enabling the model to be
trained on standard computational hardware without requiring specialized
infrastructure.

Despite these constraints, the \resnettransformer\ model demonstrated robust
performance in solar flare prediction, highlighting the effectiveness of our
method even with a relatively small and specialized dataset. The strategic
combination of efficient data processing, transfer learning by reusing a
pre-trained \resnet\ model, and optimized architectural choices ensured that our
approach remained viable and effective, delivering high predictive accuracy
without overwhelming computational demands.

Looking ahead, our methodology can be extended to develop a near-real-time
prediction system with several key enhancements. Automating the entire data
ingestion, processing, and prediction pipeline would enable continuous
real-time solar flare forecasting, providing timely alerts for impending solar
activity. This automation is essential for operationalizing our predictive model
and integrating it into space weather monitoring systems, ensuring that
predictions are delivered promptly and efficiently.

\section{Conclusion}
\label{sec:conclusion}

This study presents a novel approach for predicting multiclass solar flares by
integrating a fine-tuned convolutional neural network (\resnet) with a
Transformer-based temporal modeling framework. By leveraging domain-specific
spatial features extracted from solar images and capturing temporal
dependencies through the Transformer architecture, our method achieved superior
performance in predicting the occurrence, classification, and probability of
solar flares within 24\,hr, 36\,hr, and 72\,hr windows. This image-based,
multiclass model outperformed traditional machine learning models, including
SVMs and standalone \resnet\ CNNs, across key performance metrics such as
precision, recall, the $F_1$ score, and ROC--AUC. Additional metrics, including
balanced accuracy, the Matthews correlation coefficient, and Cohen's $\kappa$,
were introduced to evaluate the overall performance of the multiclass
classifier, with promising results. The \resnettransformer\ model underscores
the potential of combining advanced neural network architectures with
sophisticated temporal sequence modeling to enhance predictive capabilities in
solar physics.

\begin{acknowledgments}
This work utilized data downloaded from the Joint Science Operations Center
(JSOC; \url{http://jsoc.stanford.edu}), and the data were acquired by SDO/HMI.
SDO is a NASA mission, and HMI is an instrument developed by Stanford
University. The GOES X-ray flare catalogs are provided by the National Oceanic
and Atmospheric Administration. This research made use of SunPy, an
open-source and free community-developed solar data analysis package for
Python.
\end{acknowledgments}

\facilities{SDO(HMI), GOES}

\software{
  \textsf{SunPy} \citep{sunpy2020},
  \textsf{drms} \citep{glogowski2019},
  \textsf{PyTorch},
  \textsf{scikit-learn},
  \textsf{NumPy},
  \textsf{Matplotlib},
  \textsf{Astropy}
}

\bibliographystyle{aasjournalv7}
\bibliography{refs}

\end{document}